\documentclass[12pt]{article}
\usepackage{amsmath,amssymb,bm,amsthm}
\usepackage[mathscr]{eucal}

\newcommand{\SOp}[1]{\hat{#1}}
\newcommand{\hvert}{\SOp{\vphantom{0}\smash{|}}}

\newcommand{\Op}[1]{\hat{#1}}
\newcommand{\suN}{\bm{n}}
\newcommand{\suq}{\mathscr{S}}
\newcommand{\Suq}{\Op{\suq}_{^\pm}}

\newcommand{\CC}{\mathbb{C}}
\newcommand{\RR}{\mathbb{R}}
\newcommand{\ZZ}{\mathbb{Z}}

\newcommand{\ket}[1]{|#1\rangle}
\newcommand{\hVert}{\SOp{\big|}}
\newcommand{\SKet}[1]{\hVert\!#1\bigr\rangle}
\newcommand{\sket}[1]{\hvert#1\rangle}
\newcommand{\bra}[1]{\langle #1 |}

\newcommand{\sbra}[1]{\langle #1 \hvert}
\newcommand{\braket}[2]{\langle #1 | #2 \rangle}
\newcommand{\sbraket}[2]{\langle #1 \hvert #2 \rangle}
\newcommand{\SBraket}[2]{\bigl\langle #1 \hVert #2 \bigr\rangle}
\newcommand{\Eq}[1]{Eq.~(\ref{#1})}
\newcommand{\Eqs}[1]{Eqs.~({#1})}
\newcommand{\Sec}[1]{Sec.~\ref{Sec:#1}}
\newcommand{\clf}[1]{\mathfrak{#1}}

\newcommand{\Id}{\mathbf 1} 
\newcommand{\Nl}{\mathbf 0} 
\newcommand{\e}{\clf{e}}
\newcommand{\an}{\clf{a}}

\newcommand{\mi}{\mathrm i}

\newcommand{\Cl}{{\mathcal{C}\!\ell}}

\newcommand{\vac}{\varnothing}
\newcommand{\st}[2]{\stackrel{#2}{#1\mathstrut}}
\newcommand{\SKst}[2]{\SKet{\st{#1}{#2}}} 
\newcommand{\LI}{\ell}
\newcommand{\Lv}{\bm\LI}
\newcommand{\Lvac}{\Lv_\vac}
\newcommand{\Ix}{\mathcal I}
\newcommand{\IIx}{\mathit{2}\mathcal I}
\newcommand{\vx}{\bm{x}}
\newcommand{\smatr}[1]{\left(\begin{smallmatrix}#1\end{smallmatrix}\right)}
\newcommand{\per}{\pi}
\newcommand{\oper}{\Op{\per}}
\newcommand{\pmer}{\Op{\per}_{^\pm}}
\newcommand{\equ}{\simeq}
\newcommand{\nodag}{{\phantom{\dag}}}
\newcommand{\Lrho}{{\bm{\wp}}}

\newcommand{\sig}[1]{\varsigma_{#1}}

\newcommand{\sox}{\mathbin{\SOp{\otimes}}} 
\newcommand{\il}{\text{\i}_l}

\DeclareMathOperator{\Sc}{Sc}

\title{\bf Jordan-Wigner transformation and\\ qubits with nontrivial exchange rule}
\date{}
\author{{\em Alexander Yu.\ Vlasov}}

\begin{document}
	
\maketitle

\begin{abstract}
  Well-known (spinless) fermionic qubits may need more subtle consideration in comparison with
  usual (spinful) fermions. Taking into account a model with local fermionic modes, formally only the 
  `occupied' states $|1\rangle$ could be relevant for antisymmetry with respect to particles interchange, 
  but `vacuum' state $|0\rangle$ is not. Introduction of exchange rule for such fermionic qubits 
  indexed by some   `positions' may look questionable due to general super-selection principle.
  However, a consistent algebraic construction of such `super-indexed' qubits is presented in this work.
  Considered method has some relation with construction of super-spaces, but it has some differences
  with standard definition of supersymmety sometimes used for generalizations of qubit model.
\end{abstract}

\section{Introduction}
\label{Sec:Intro}

Analogues of fermionic creation and annihilation (ladder) operators
were suggested by Richard Feynman for description
of quantum computers already in the very first works \cite{FeySim,FeyComp}.
However, Jordan-Wigner transformation \cite{JW} is necessary 
to make such operators anticommutting for 
{\em different} qubits.
Such approach was used later in so-called {\em fermionic quantum
computation} \cite{BK00}. 

Representation of fermionic ladder operators in such a way formally requires 
some consequent indexing (order) for description of Jordan-Wigner transformation.
The order does not manifest itself directly in algebraic properties
of ladder operators, but transformations of states formally depend on
such indexes in rather nonlocal way.   

States of physical bosons and fermions can be described in natural way by 
symmetric and antisymmetric tensors respectively, but fermionic quantum computation
is rather relevant with more subtle exchange behavior of some quasi-particles.

Formally, qubits in state $\ket{0}$ corresponds to `empty modes' 
and only qubits in state $\ket{1}$ treated as `occupied modes' 
could be relevant to fermionic exchange principle for qubits
marked by some indexes.
Consistent mathematical model
of such `super-indexed' states is suggested in presented work.

\medskip

More detailed description of such states is introduced in \Sec{SigStat}
together with formal definition of {\em signed exchange rule} and `super-indexed' 
qubits denoted further as {\em $\suq$-qubits}. 
The different kinds of operators acting on $\suq$-qubits are constructed 
in \Sec{SigOp}.

Algebraic models of $\suq$-qubits are discussed in \Sec{Mod}. The 
`non-trivial' non-commutative part of such model is similar
with exterior algebra recollected in \Sec{Ext}. However, `trivial'
commutative elements could not be naturally presented in such a way
and more complete model is suggested in \Sec{Clif}. The Clifford
algebras initially used in \Sec{SigOp} for applications to 
gates and operators become the basic tools here. The $\suq$-qubits
are introduced as minimal left ideals of the Cliford algebras.
Finally, some comparison with possible alternative models of qubits
related with super-spaces are outlined in \Sec{Supvec}.

\section{Super-indexed qubits}
\label{Sec:Sign}

\begin{flushright}
	\raisebox{-.5ex}[1.2ex][0ex]{\LARGE$\mathfrak{F}$}%
	$\mathfrak{lesh\ flourisht\ of\ fermison}$~~\\ 
	$\mathfrak{with\ frumentee\ noble.}$
	\footnotesize
	
	(Alliterative {\sl Morte Arthure})
\end{flushright}

\subsection{States}
\label{Sec:SigStat}

Let us introduce special notation for qubits marked by some set of indexes $\Ix$
with basic states denoted as
\begin{equation}\label{suq}
\SKst{\mu}a \SKst{\nu}b  \cdots 
=\SKet{\st{\mu}a,\st{\nu}b, \dots },
\quad a, b, \ldots  \in \Ix, \quad \mu,\nu,\ldots \in \{0,1\} .
\end{equation}
All indexes in the sequence $a,b,\ldots$ must be {\em different}.
The $\Ix$ can be associated with some nodes in 
multi-dimensional lattices, more general graphs or other configurations 
{\em without natural ordering}.
Thus, the qubits in \Eq{suq} may be rearranged in different ways.

An idea about basic states of qubits as `occupation numbers' 
of anti-commuting `local fermionic modes' (LFM) \cite{BK00} can be
formalized by introduction of equivalence relation between elements \Eq{suq} 
with different order of the indexes defined for any neighboring pair by
{\em signed exchange rule} 
$$
\SKst0a \SKst0b  \equ \SKst0b\SKst0a ,\quad \SKst0a \SKst1b  \equ \SKst1b\SKst0a ,\quad
\SKst1a \SKst1b  \equ - \SKst1b\SKst1a,
$$
{\em i.e.},
\begin{equation}\label{supcom}
	\SKst{\mu}a \SKst{\nu}b \equ (-1)^{\mu\nu}\SKst{\nu}b\SKst{\mu}a , 
	\quad \forall a\neq b \in \Ix, \quad \mu,\nu \in \{0,1\}.
\end{equation}

Due to such rules terms $\sket{0}$ with `attached' indexes can be exchanged (`commute')
with any state $\sket{\psi}=\alpha\sket{0}+\beta\sket{1}$, but two $\sket{1}$ 
require change of the sign for such a swap (`anti-commute'). 
For standard notation and qubits without special indexes the exchange rule could be 
implemented by {\em signed swap operator}
\begin{equation}\label{sswap}
\Suq =
\smatr{
	1 & ~0 & ~0 & ~0\\
	0 & ~0 & ~1 & ~0\\
	0 & ~1 & ~0 & ~0\\
	0 & ~0 & ~0 & -1
}\!.
\end{equation}
States \Eq{suq} with equivalence relation \Eq{supcom} define basis of
some linear space $\suq$ denoted here as {\em `super-indexed' qubits}
or {\em $\suq$-qubits}.

The equivalence relation \Eq{supcom} can be extended on arbitrary permutation $\per$.
Such operator is denoted further $\pmer$ and notation $\oper$ is reserved for usual permutation.
The construction of $\pmer$ does not depend on decomposition of $\per$ on adjacent transpositions,
{\em i.e.}, swaps of $\suq$-qubits considered above. Such consistency becomes more natural 
from algebraic models below in \Sec{Mod} and `physical' interpretation with LFM.

It can be also proved for arbitrary state by direct check for the basis.
Let us consider for a given basic state different sequences of transpositions 
produced the {\em same permutation}. It is necessary to show that the sign does not depend 
on the decomposition of the permutation into the sequence. Let us consider restriction 
($\per_1$) of permutation on subset of indexes corresponding to $\suq$-qubits with unit value. 
For the only nontrivial case the restriction of swap on such subset corresponds 
to exchange of two units with change of sign. 
So, for any decomposition the basic vector may change the sign only if the
permutation $\per_1$ is odd. Thus, $\pmer$ is the same for any decomposition of $\oper$
on transpositions defined by {\em signed exchange rule} \Eq{supcom}. 

\smallskip

The relation $\Suq$ can be considered as a formalization of swap with two LFM denoted as `$\Leftrightarrow$' 
in \cite{BK00}. It could be expressed as composition of usual exchange of qubits `$\leftrightarrow$' and 
{\em `swap defect' operator} \cite{BK00} 
\begin{equation}\label{dswap}
\Op{D} =
\smatr{
	1 & ~0 & ~0 & ~0\\
	0 & ~1 & ~0 & ~0\\
	0 & ~0 & ~1 & ~0\\
	0 & ~0 & ~0 & -1
}\!.
\end{equation}

Perhaps, the term `fermionic qubits' might be
not very justified for the model considered here, because
the property \Eq{supcom} would correspond to fermion for $\sket1$ 
(`occupied,' $n=1$) and boson for $\sket0$ (`empty,' $n=0$).

Thus, $\suq$-qubits could be considered as quasi-particles (`fermisons')
with {\em combined statistics}, because exchange rule instead of $(\pm 1)$ 
multiplier for bosons or fermions should use {\em swap defect} operator  
\begin{equation}\label{supsup}
\SKst{\psi}a \SKst{\phi}b  \mapsto (-1)^{\Op{n}_a\Op{n}_b}\Bigl(\SKst{\phi}b\SKst{\psi}a\Bigr), 
\end{equation} 
where a formal representation $\Op{D} = (-1)^{\Op{n}_a\Op{n}_b}$ is used,   
where $\Op{n}_a$ and $\Op{n}_b$ are analogues of {\em occupation number operators} defined for 
usual qubit as 
\begin{equation}\label{ocnum}
\Op{n}\ket{\nu} = \nu\ket{\nu}, \quad \nu\in\{0,1\},\qquad
\Op{n} = \begin{pmatrix}
0 & ~0\\
0 & ~1 
\end{pmatrix}\!.
\end{equation} 
The result of a swap \Eq{supsup} 
is defined in simple `product' form \Eq{supcom} for the basis,
but for arbitrary states the expressions are less trivial. 

\smallskip

The $\suq$-qubit also could be compared with an element of {\em super vector space}, 
but due to some subtleties outlined in \Sec{Supvec} such approach should be discussed
elsewhere.

\smallskip

The scalar product of $\suq$-qubits states can be naturally defined for 
the equivalent sequences of indexes $S$ in both (`bra' and `ket') parts
\begin{equation}\label{suqscal}
\SBraket{\st \Psi S}{\st \Phi S} = \braket{\Psi}{\Phi}. 
\end{equation}
The definition \Eq{suqscal} does not depend on a sequence $S$. Indeed, let us consider
permutations of indexes $\per\colon S \mapsto S'$. For the basic states a permutation
may only introduce $(\pm 1)$ multiplier and the scalar product \Eq{suqscal} does not change. 
It can be also checked directly for arbitrary states 
\begin{equation}\label{permscal}
\SBraket{\st{\Psi'}{S'}}{\st{\Phi'}{S'}} = \bra{\Psi}\pmer^\dag\,\pmer^\nodag\ket{\Phi}
=\braket{\Psi}{\Phi} = \SBraket{\st \Psi S}{\st \Phi S}.
\end{equation}

\subsection{Operators}
\label{Sec:SigOp}

Description of quantum gates with {\em  annihilation and creation} (`ladder') operators 
was initially suggested by R.\ Feynman \cite{FeySim,FeyComp}. However, despite of  
formal resemblance with Pauli exclusion principle for fermions
\begin{equation}\label{ladloc} 
\Op a^\dag\ket{0} = \ket{1},\quad\Op a \ket{1} = \ket{0},\quad\Op a \ket{0} = \Op a^\dag \ket{1} = 0,
\tag{$a$}
\end{equation}
they do not satisfy {\em canonical anticommutation relation} 
(CAR) for different qubits. 
Sometimes, usual qubits are compared with so-called `hardcore' bosons,
but it is not discussed here. 
It is considered instead, how ladder operators with CAR 
can be introduced for $\suq$-qubits, see \Eq{CAR} below.  

The {\em creation operator} $\an^\dag_a$  
can be defined for basic states taking into account exchange rule \Eq{supcom} of {$\suq$-qubits}
\begin{equation}\label{supcre}
\an^\dag_a \SKet{\,\st\dots{L}, \st 0 a, \st\dots{R}} 
= \SKet{\st 1 a, \st\dots{L}, \st\dots{R}},
\quad \an^\dag_a\SKet{\,\st\dots{L}, \st 1 a, \st\dots{R}} = 0,
\end{equation}
where $L$ and $R$ correspond to arbitrary sequences before and after index 
`$a$' respectively. 
The conjugated {\em annihilation operator} $\an_a= (\an_a^\dag)^\dag$ 
in simpler case with index `$a$' in the first position 
can be written
\begin{equation}\label{supann1}
\an_a\SKet{\st0a, \dots} = 0,\quad 
\an_a\SKet{\st1a, \dots} = \SKet{\st0a, \dots}.
\end{equation}
A sign for application of operator $\an_a$ to arbitrary position should be found 
using rearrangement of indexes together with signed exchange rule \Eq{supcom}.

Let us denote $\pm_a$ a sign derived from \Eq{supcom} for 
expressions such as
\begin{equation}\label{pma}
\SKet{\,\st\dots{L}, \st 1 a, \st\dots{R} }  
=  \pm_a \SKet{\st 1 a, \st\dots{L}, \st\dots{R} },
\end{equation}
where $\pm_a = (-1)^{\#L}$ with $\#L=\sum_{l \in L}n_l$ is number of units 
in sequence $L$ of positions before `$a$'. for simplicity $L$ and $R$ 
are omitted further. 

Finally, \Eqs{\ref{supcre},\ref{supann1}} can be rewritten
\begin{subequations}\label{supca}
\begin{align}\label{supcrea}
&\an_a^\dag\SKet{\,\dots,\st 0 a, \dots} = \pm_a\SKet{\,\dots,\st 1 a, \dots},&
&\an_a^\dag\SKet{\,\dots, \st 1 a, \dots} = 0,
\\ \label{supann}
&\an_a\SKet{\,\dots, \st 0 a, \dots} = 0,&
&\an_a\SKet{\,\dots, \st 1 a, \dots} = \pm_a\SKet{\,\dots,\st 0 a, \dots}.
\end{align}
\end{subequations}

For consequent indexes $a=0,\dots,m-1$ \Eq{supca} is in agreement with usual
Jordan-Wigner transformation \cite{JW,BK00}, {\em i.e.,}
\begin{subequations}\label{JWac}
\begin{align}
\an_a \ket{n_0,\ldots,n_{a-1},1,n_{a+1},\ldots} &=
(-1)^{\sum_{k=0}^{a-1}n_k}\ket{n_0,\ldots,n_{a-1},0,n_{a+1},\ldots}\notag\\
\an_a \ket{n_0,\ldots,n_{a-1},0,n_{a+1},\ldots} &= 0, 
\end{align}
and $\an_a^\dag$ is Hermitian conjugation
\begin{align}
	\an_a^\dag \ket{n_0,\ldots,n_{a-1},0,n_{a+1},\ldots} &=
	(-1)^{\sum_{k=0}^{a-1}n_k}\ket{n_0,\ldots,n_{a-1},1,n_{a+1},\ldots}\notag\\
	\an_a^\dag \ket{n_0,\ldots,n_{a-1},1,n_{a+1},\ldots} &= 0. 
\end{align}
\end{subequations}

The $\an_a$ and $\an_a^\dag$ defined in such a way
satisfy {\em canonical anticommutation relations} 
\begin{equation}\label{CAR}
 \{\an_a,\an_b\} = \{\an_a^\dag,\an_b^\dag\} = \Nl,
 \quad \{\an_a,\an_b^\dag\} = \delta_{ab}\Id.
\end{equation}

\medskip

Let us now introduce {\em Clifford algebra} $\Cl(2m)$ with $2m$ generators 
using operators \Eq{JWac}
\begin{equation}\label{a2e}
\e_a = \mi(\an^\dag_a + \an_a),\quad
\e'_a = \an_a - \an^\dag_a.
\end{equation}
Operators of so-called {\em Majorana fermionic modes\/}  coincides 
with \Eq{a2e} up to imaginary unit multiplier \cite{BK00}.

With earlier definitions of annihilation and creation operators it
can be expressed for basis as
\begin{align}\label{supgen}
&\e_a\SKet{\dots, \st 0 a, \dots}  = \pm_a\mi\, \SKet{\dots,\st 1 a, \dots},& 
&\e_a\SKet{\dots,\st 1 a, \dots}   = \pm_a\mi\, \SKet{\dots,\st 0 a, \dots}, \notag\\ 
&\e'_a\SKet{\dots,\st 0 a, \dots}  = \mp_a\, \SKet{\dots,\st 1 a, \dots},&
&\e'_a\SKet{\dots,\st 1 a, \dots}  = \pm_a\, \SKet{\dots,\st 0 a, \dots},
\end{align}
where $\mp_a = - (\pm_a)$.

\smallskip

For consequent indexes $j=0,\dots,m-1$ the \Eq{supgen} again correspond to Jordan-Wigner 
formalism with definition of complex Clifford algebra $\Cl(2m,\CC)$ by tensor product of Pauli matrices
\cite{JW,ClDir}
\begin{eqnarray}\label{clgen}
	\e_{j} & = &
	\mi\,{\underbrace{\Op\sigma^z\otimes\cdots\otimes \Op\sigma^z}_{j}\,}\otimes
	\Op\sigma^x\otimes\underbrace{\Op\Id\otimes\cdots\otimes\Op\Id}_{m-j-1} \, ,
	\notag\\
	\e'_{j} & = &
	\mi\,{\underbrace{\Op\sigma^z\otimes\cdots\otimes \Op\sigma^z}_{j}\,}\otimes
	\Op\sigma^y\otimes\underbrace{\Op\Id\otimes\cdots\otimes\Op\Id}_{m-j-1} \, .
\end{eqnarray}

However, \Eq{clgen} directly introduces order of indexes unlike more abstract definitions of 
operators such as \Eq{supca} and \Eq{supgen} respecting structure 
of $\suq$-qubits without necessity of predefined order.  

\smallskip

The linear combinations of all possible products with operators $\e_a$,  $\e'_a$
(or $\an_a$, $\an_a^\dag$) 
for given set  $\Ix$ with $m_\Ix$ indexes generate Clifford algebra $\Cl(2m_{\Ix},\CC)$ with 
dimension $2^{2m_\Ix}$.
Thus, an arbitrary linear operator on space $\suq$ can be represented in such a way,
but unitarity should be also taken into account for construction of quantum gates
on $\suq$-qubits.

An alternative notation $\e_{a'}=\e'_a$, $a' \in \Ix' \sim \Ix$ unifying two sets of generators 
from \Eq{a2e} into the single collection with doubled set of indexes 
$\IIx = \Ix \cup \Ix'$ is also used further for brevity.
Definition of Clifford algebra $\Cl(2m_{\Ix},\CC)$ can be written with such a set as
\begin{equation}\label{ecom}
\{\e_a,\e_b\} = -2 \delta_{ab}\Id, \quad a,b \in \IIx.
\end{equation} 
and conjugation of elements as  
\begin{equation}\label{edag}
 \e_a^\dag = -\e_a^\nodag, \quad a \in \IIx .
\end{equation}

The elements \Eq{a2e} generate $\Cl(2m_{\Ix},\CC)$ isomorphic with whole 
algebra of \mbox{$2^m \times 2^m$} complex matrices. The unitary gates may be expressed
as exponents of Hermitian elements with pure imaginary multipliers discussed below.

Let us consider for some sequence $L$ with $l$ indexes from $\IIx$ 
 a product of $l$ generators   
\begin{equation}\label{prode}
 \e_L = \e_{a_1} \cdots \e_{a_l},\quad a_1,\ldots,a_l \in \IIx\!.
\end{equation}
Linear subspaces $\Cl^{(l)}$ is introduced as a span of such products, $\e_L\in\Cl^{(l)}$. 
The notation $\Cl^0$ and $\Cl^1$ is reserved here for standard decomposition 
of $\Cl$ as $\ZZ_2$-graded algebra 
with two subspaces corresponding to linear span of all possible products with
even and odd $l$ respectively \cite{Post} 
\begin{equation}\label{gradCl}
  \Cl(n) = \Cl^0(n)\oplus\Cl^1(n). 
\end{equation}
The square of element $\e_L$ can be expressed as
\begin{equation}\label{eLq}
 \e_L^2 = (-1)^{\sig l}, \quad \sig l =  \frac{l\,(l+1)}{2} \bmod 2.
\end{equation}
All such elements are unitary with
respect to conjugation operation \cite{ClDir}
\begin{equation}\label{eLdag}
\e_L^\dag = (-1)^{\sig l}\e_L^\nodag,\qquad
\e_L^\dag \e_L^\nodag = \Id. 
\end{equation}
The construction of Hermitian basis is also straightforward 
\begin{equation}\label{eLHerm}
(\mi^{\sig l}\e_L)^\dag = (-\mi)^{\sig l}\e_L^\dag  =
(-\mi)^{\sig l}(-1)^{\sig l}\e_L^\nodag = \mi^{\sig l}\e_L.
\end{equation}

An exponential representation of unitary operators is simply derived from
\Eq{eLHerm} for arbitrary compositions of basic elements, {\em e.g.},
for  $\clf h_l \in \Cl^{(l)}$ 
\begin{equation}\label{expum}
 \clf u(\tau) = \exp(-\il \clf h_l \tau),\quad \il = \mi \cdot \mi^{\sig l} = \mi^{\sig l + 1}, \quad \tau \in \RR.
\end{equation}
Due to property $\sig{l+4}=\sig l$ multipliers can be given in the table,   
\begin{equation}\label{sqtab}
\begin{array}{|c|r|r|r|r|}
\hline	
l \bmod 4&0&1&2&3\\ \hline
\sig l&0&1&1&0\\ \hline 
-\il&-\mi&\ 1&\ 1&-\mi\\ \hline
\end{array}
\end{equation}

\smallskip

Expression of unitary group $U(2^m)$ using families of quantum gates 
can be derived using approach with exponents due to correspondence
between Lie algebras and Lie groups. The method initially was suggested for 
construction of universal set of quantum gates \cite{Div95,DBE95,Vla0}. 
Clifford algebra $\Cl(2m)$ with {\em Lie bracket operation} defined as 
a standard commutator
\begin{equation}\label{com}
[\clf{a,b}] = \clf{a b - b a} 
\end{equation}
can be used for representation of  Lie algebra of special unitary group $su(2^m)$ and group 
$SU(2^m)$ can be expressed as exponents of elements from $\Cl(2m)$. 

\medskip 

In the exponential representation analogue of one-gates for $\suq$-qubits with $l=1,2$ can be 
expressed as 
\begin{equation}\label{expu}
 \clf u_j = \exp(h_1 \e_j + h_2 \e'_j + h_3 \e_j\e'_j), \quad h_1,h_2,h_3 \in \RR.
\end{equation}
It can be also rewritten
\begin{equation}\label{quatu}
\clf u_j = q_0 + q_1 \e_j + q_2 \e'_j + q_3 \e_j\e'_j, \quad 
q_0^2+q_1^2+q_2^2+q_3^2=1,~
q_0,q_1,q_2,q_3 \in \RR.
\end{equation}

Analogue of two-gates for $\suq$-qubits with indexes $j, k \in \Ix$ can be declared
by analogue exponents with linear combination of different 
products including $\e_j$, $\e'_j$, $\e_k$, $\e'_k$ with coefficients 
are either real ($l=1,2$) or pure imaginary ($l = 3,4$).

Similar exponents with more general elements from linear subspaces $\Cl^{(l)}$ for
$l=2$ and $l=1,2$ (with arbitrary combinations of indexes from $\IIx$) 
generate `non-universal' subgroups isomorphic to Spin$(2m)$ and Spin$(2m+1)$ respectively 
\cite{ClDir,Vla0}, but inclusion element with $l=3$ is enough to generate  
unitary group $SU(2^m)$ \cite{Vla0}.

\smallskip

Due to physical reasons for some models only terms with 
{\em even number} of generators 
must be used \cite{BK00}. 
Formally, such terms belong to even subalgebra $\Cl^0$ that may be
again treated as a Clifford algebra due to standard isomorphism 
\mbox{$\Cl(n-1) \cong \Cl^0(n)$} \cite{Post}
\begin{equation}\label{evisom}
 \Cl(n-1) \to \Cl^0(n), \qquad
 \e_L \mapsto 
 \begin{cases}  
  \e_L,& \e_L \in \Cl^0(n-1),\\
  \e_L \e_{n},& \e_L \in \Cl^1(n-1).
 \end{cases}
\end{equation}
Thus, a model with even number of generators in Hamiltonians 
and universal subset of quantum gate with $l=2,4$ \cite{BK00} 
is also described by Clifford algebra due to isomorphism $\Cl^0(2m) \cong \Cl(2m-1)$.

\section{Algebraic models of $\bm\suq$-qubits}
\label{Sec:Mod}

\subsection{Exterior algebra}
\label{Sec:Ext}

Let us consider a vector space $V$ with basis $\vx_j$, 
$j = 0,\ldots,m-1$.
The exterior (Grassmann) algebra is defined as
\begin{equation}\label{Lbas}
\Lambda(V) = \bigoplus_{k=0}^m \Lambda^k(V),
\end{equation}
where $\Lambda^0(V)$ are scalars, $\Lambda^1(V)=V$ are vectors,
and $\Lambda^k(V)$, $k>1$ are antisymmetric $k$-forms (tensors) with basis
\begin{equation}\label{Lkbas}
 \vx_{j_1} \wedge \cdots \wedge \vx_{j_k}, \quad j_1 < \cdots < j_k
\end{equation}
where `$\wedge$' denotes antisymmetric (exterior) 
product $\vx_j \wedge \vx_k = - \vx_k \wedge \vx_j$, 
$\vx \wedge \vx = \Nl$, $\forall \vx \in V$.

The dimension of whole space $\Lambda(V)$ is $2^m$ and any basic state
\Eq{suq} of {$\suq$-qubits} could be mapped into $\Lambda(V)$
\begin{equation}\label{suqlam}
 \SKet{\st{n_{j_1}}{j_1},\dots,\st{n_{j_m}}{j_m}} \mapsto
 \bigwedge_{\substack{j \in \Ix \\n_{j} = 1}} \vx_{j}.
\end{equation}
Such a method inserts into exterior product only $\vx_j$ 
with indexes $j$ satisfying $n_j=1$. However,
\Eq{suqlam} is one-to-one map and arbitrary form $\Omega \in \Lambda(V)$ 
corresponds to some state $\sket{\Omega}$ up to appropriate normalization.

The creation and annihilation operators in such representation
correspond to a known construction of Clifford algebra using
space of linear transformations on $\Lambda(V)$ \cite{ClDir}
and may be expressed for basis \Eq{Lkbas}
\begin{subequations}\label{caLam}
\begin{align}
 \an_j^\dag &\colon \vx_{j_1} \wedge \cdots \wedge \vx_{j_k} &\mapsto & 
  \quad \vx_j \wedge \vx_{j_1} \wedge \cdots \wedge \vx_{j_k}, 
  \label{creLam}\\
 \an_j  &\colon \vx_{j_1} \wedge \cdots \wedge \vx_{j_k} &\mapsto &
 \quad \sum_{l=1}^k(-1)^l \vx_{j_1} \wedge \cdots \wedge (\delta_{j, j_l}\Id)\wedge \cdots \wedge \vx_{j_k}, 
 \label{anLam}
\end{align}
\end{subequations}
where $\Id$ is unit of algebra $\Lambda(V)$ and notation 
$\Id \wedge \omega = \omega \wedge\Id = \omega$, $\omega \in \Lambda(V)$ 
is supposed in \Eq{caLam}.
Such operators satisfy \Eq{CAR} and respect map \Eq{Lkbas}
due to consistency of \Eq{creLam} with \Eq{supcre} and \Eq{anLam} with \Eq{supann}.

The generators of complex Clifford algebra $\Cl(2m,\CC)$ can be expressed with
earlier defined {\em pair} of generators \Eq{a2e} for each index and for real case 
elements $\e'_j$ might be used to produce $\Cl(m,\RR)$ \cite{ClDir}.

The considered representation of {$\suq$-qubits} with exterior algebra
$\Lambda(V)$ despite of one-to-one correspondence \Eq{suqlam} for complete basis 
may be not very convenient for work with `reduced' expressions
such as \Eq{supcom}, because only qubits with state $\sket{1}$ map into different $\vx_a$, 
but any sequence of qubits in state $\sket{0}$ formally corresponds 
to unit scalar $\Id \in \Lambda^0(V)$. 
An approach with Clifford algebras discussed next helps to avoid such a problem.

\subsection{Clifford algebras and spinors}
\label{Sec:Clif}

For Clifford algebra $\Cl = \Cl(2m,\CC)$ the space of spinors has 
dimension $2^m$ and it can be represented as {\em minimal left ideal} \cite{Lou}.
The left ideal $\LI \subset \Cl$ by definition has a property 
\begin{equation}\label{LId}
 \clf c \, \clf l  \in \LI\colon \quad  
 \forall \clf l \in \LI ,~ \clf c \in \Cl. 
\end{equation}
By definition, the (nonzero) {\em minimal} left ideal does not contain any other (nonzero) left ideal.

The notation with single set of indexes $a \in \Ix$ and $m$ pairs
of generators $\e_a$ and $\e_a'$ is again used below.
Annihilations and creation operators corresponding \Eq{a2e}
are also useful further
\begin{equation}\label{e2a}
 \an_a = \frac{\e_a + \mi \e'_a}{2\mi}, \quad
 \an_a^\dag = \frac{\e_a - \mi \e'_a}{2\mi}.
\end{equation}

The minimal left ideal $\LI$ is generated by all possible 
products with an appropriate element $\Lvac$
\begin{equation}\label{lidvac}
 \LI = \Bigl\{\clf c \Lvac \colon \clf c \in \Cl, ~
 \Lvac = \prod_{a \in \Ix} \Lv_0^a \Bigr\}
\end{equation} 
where
\begin{equation}\label{lv0}
\Lv_0^a = \frac{\Id + \mi \e^{}_a \e'_a}{2} = \an^{}_a \an_a^\dag
\end{equation}
are $N$ commuting projectors $(\Lv_0^a)^2 = \Lv_0^a$. 
Due to identity $\Lv_0^a = \mi \e^{}_a \e'_a\Lv_0^a$
for any index $a$ it can be written 
\begin{equation}\label{evac}
\e'_a \Lvac = -\mi\e_a\Lvac. 
\end{equation}

Let us apply definition of $\LI$ \Eq{lidvac} to linear decomposition 
of $\clf c$ on terms with products of generators $\e_a$ and $\e_a'$.
Any element of $\LI$ in \Eq{lidvac} may be rewritten
as a linear combination of terms without $\e'_a$ due to \Eq{evac}.
Thus, $\LI$ has dimension $2^m$ with products at most $m$
different generators $\e_a$ on $\Lvac$ as a basis.
Let us also introduce notation 
\begin{equation}\label{lv1}
\Lv_1^a  = \an_a^\dag \Lv_0^a 
 = \an_a^\dag 
\end{equation}
then a basis of spinor space $\LI$ can be rewritten in agreement with \Eq{suq}
\begin{equation}\label{LIbas}
 (\Lv_\mu^a \Lv_\nu^b \cdots ) \Lvac \longleftrightarrow
 \SKet{\st{\mu}a,\st{\nu}b, \ldots },
 \quad a, b, \ldots  \in \Ix, \quad \mu,\nu,\ldots \in \{0,1\}, 
\end{equation}
there all indexes $a, b, \ldots$ {\em must be different}.

In representation \Eq{clgen} with consequent indexes $a = 0,\dots,m-1$ 
the elements $\Lv_0^a$ correspond to $2^m \times 2^m$ diagonal matrices with 
units and zeros described by equation
\begin{equation}\label{Lv0a}
\Lv_0^a  \mapsto {\underbrace{\Id\otimes\cdots\otimes\Id}_{a}\,}\otimes
\Lv_0\otimes\underbrace{\Id\otimes\cdots\otimes\Id}_{m-a-1} \, ,
\end{equation}
where $\Lv_0 = \ket{0}\bra{0}$.
Therefore, $\Lvac$ corresponds to a $2^m \times 2^m$ diagonal matrix with unit only in the 
very first position
\begin{equation}\label{mlvac}
\Lvac \mapsto \underbrace{\Lv_0\otimes\cdots\otimes\Lv_0}_m
=\ket{\underbrace{0,\dots,0}_m}\bra{\underbrace{0,\dots,0}_m}.
\end{equation}
In such a case a product $\clf c \Lvac$ in definition of ideal $\LI$ \Eq{lidvac} 
corresponds to a $2^m \times 2^m$ matrix with only nonzero first column.
It can be used for representation of a vector with $2^m$ components.

For arbitrary sequences of indexes from a set $\Ix$ an analogue of \Eq{supcom} also holds
\begin{equation}\label{LIcom}
 \Lv_\mu^a \Lv_\nu^b = (-1)^{\mu\nu}\Lv_\nu^b\Lv_\mu^a,
 \quad a \neq b \in \Ix, \quad \mu,\nu \in \{0,1\}. 
\end{equation}
The {\em inequality} of indexes $a \neq b$ is essential, because the
{\em super-commutativity} \Eq{LIcom} 
{\em does not} hold for $a = b$ if $\mu \neq \nu$
\begin{equation}\label{L0L1}
 \Lv_0^a\Lv_0^a = \Lv_0^a,\quad
 \Lv_1^a\Lv_1^a = \Lv_0^a \Lv_1^a = \Nl, \quad
 \Lv_1^a \Lv_0^a = \Lv_1^a, \quad
 \Lv_0^a \Lv_1^a \neq \Lv_1^a \Lv_0^a.
\end{equation}
Anyway, all indexes of $\suq$-qubits \Eq{suq} are different by definition   
and inequality in \Eq{L0L1} does not affect considered representation \Eq{LIbas}.

\smallskip

However, some other expressions for operators or scalar product \Eq{suqscal}  
may require more general combinations of indexes. It is discussed below.

For arbitrary element $\Lv \in \LI$ 
the operators $\an_a$, $\an^\dag_a$ can be naturally defined via left multiplication
$$
 \an_a \colon \Lv \mapsto \an_a \Lv, \quad \an_a^\dag \colon \Lv \mapsto \an_a^\dag \Lv.
$$
With respect to map \Eq{LIbas} it corresponds to \Eq{supca}. Let us check that.

Operators $\an_a$, $\an^\dag_a$ commute with $\Lv_0^b$ and anticommute with $\Lv_1^b$ for $a \neq b$.
For equivalent indexes quite natural expressions follow from definitions    
\begin{equation}\label{aLa}
  \an_a \Lv_0^a = \Nl, \quad \an_a \Lv_1^a = \Lv_0^a, \quad
  \an_a^\dag \Lv_0^a = \Lv_1^a, \quad \an_a^\dag \Lv_1^a = \Nl.
\end{equation}
 
Let us rewrite map \Eq{LIbas} with shorter notation for basic states $\suN$
\begin{equation}\label{LIbas'}
\SKet{\st{\suN}{\Ix}} = \SKet{\st{n_{j_a}}{j_a},\st{n_{j_b}}{j_b},\dots} \longleftrightarrow
\Lv_{\suN} = \prod_{j \in \Ix} \Lv_{n_j}^j.
\tag{\ref{LIbas}$'$}
\end{equation}
It may be also expressed in an alternative form
\begin{equation}\label{LIbasan}
 \Lv_{\suN} = \prod_{j \in \Ix}{(\an_j^\dag)\!^{}}^{n_j}\!\Lv_0^j =
 \Bigl(\prod_{\substack{j \in \Ix \\n_{j} = 1}}\an_j^\dag\Bigr) \Lvac.
\end{equation}
resembling \Eq{suqlam} for Grassmann algebra.
Due to \Eq{lidvac} products of operators $\an^\dag_a$ are mapped by \Eq{LIbasan}
into elements of left ideal of Clifford algebra, {\em cf\/} \Eq{LId}. 

With respect to map \Eq{LIbas} actions of $\an^\dag_a$ are in agreement with 
\Eq{supcre} and $\an_a$ satisfy an analogue of \Eq{supann1}.
Thus, operators $\an_a, \an^\dag_a$ and their linear combinations
$\e_a, \e'_a$ are corresponding to \Eq{supca} and \Eq{supgen} respectively.

\smallskip

The scalar product \Eq{suqscal} also can be naturally expressed.
Let us find conjugations of $\Lv_0^a$ and $\Lv_1^a$ 
\begin{equation}\label{lvc}
 {\Lv_0^a}^\dag = \an^{}_a \an_a^\dag = \Lv_0^a, \quad
 {\Lv_1^a}^\dag = \an_a.
\end{equation}
The equation 
\begin{equation}\label{LdagL}
{\Lv_j^a}^\dag \Lv_k^a = \delta_{jk}\Lv_0^a, \quad j,k = 0,1
\end{equation}
can be checked directly
$$
{\Lv_0^a}^\dag \Lv_0^a = {\Lv_1^a}^\dag \Lv_1^a = \Lv_0^a, \quad
{\Lv_0^a}^\dag \Lv_1^a = {\Lv_1^a}^\dag \Lv_0^a = \Nl
$$
together with appropriate expressions for products \Eq{LIbas'} 
\begin{equation}\label{ortLbas}
{\Lv^\dag_{\suN}} \Lv_{\suN}^\nodag = \Lvac, \quad
{\Lv^\dag _{\suN}}\Lv_{\suN'}^{~} = \Nl.
\end{equation}
Let us also use notation $\Lv_{\Psi}$ for representation of arbitrary $\sket{\Psi}$, 
{\em i.e.}, linear combinations of basic states. Then scalar product \Eq{suqscal}
can be written using \Eq{ortLbas}
\begin{equation}\label{Lscal}
\Lv^\dag_{\Psi} \Lv^\nodag_{\Phi} = \SBraket{\st \Psi \Ix}{\st \Phi \Ix} \Lvac = \sbraket{\Psi}{\Phi}\Lvac,
\end{equation} 
where super-index $\Ix$ denotes set of indexes used in \Eqs{\ref{LIbas}, \ref{LIbas'}} and 
it can be dropped, because all indexes are 
naturally taken into account in such algebraic expressions with
appropriate order.

Special notation can be used for `scalar part' of an element 
\begin{equation}\label{Sc}
\clf c \in \Cl(2m,\CC), \quad \clf c = c\Id + \cdots , \quad \Sc(\clf c) = c.
\end{equation} 
\Eq{Lscal} can be rewritten now to express the scalar product directly
\begin{equation}\label{ScSc}
 \Sc(\Lv^\dag_{\Psi} \Lv^\nodag_{\Phi})  = \Sc\bigl(\sbraket{\Psi}{\Phi}\Lvac\bigr)=
 \sbraket{\Psi}{\Phi}\Sc(\Lvac) = 2^{-m}\sbraket{\Psi}{\Phi}.
\end{equation}

Let us introduce an analogue of density operator. For pure state
it can be defined 
\begin{equation}\label{Ldens}
 \Lrho_\Psi = \Lv^\nodag_{\Psi} \Lv^\dag_{\Psi}  
\end{equation}
with natural property
\begin{equation}\label{Lproj}
 \Lrho_\Psi \Lv_\Phi = \Lv^\nodag_{\Psi} \Lv^\dag_{\Psi} \Lv_\Phi 
 = \Lv_{\Psi} \sbraket{\Psi}{\Phi} \Lvac = \sbraket{\Psi}{\Phi} \Lv_{\Psi}\Lvac = \sbraket{\Psi}{\Phi} \Lv_{\Psi}. 
\end{equation}
Arbitrary operators can be expressed using linear combinations with pairs of basic states
\begin{equation}\label{Lbasop}
\sket{\suN'}\sbra{\suN} \longleftrightarrow
\Lrho_{\suN'\!,\suN} = \Lv^{~}_{\suN'} \Lv^\dag_{\suN},\qquad
\Lrho_{\suN'\!,\suN} \Lv_\Phi 
= \Lv_{\suN'} \sbraket{\suN}{\Phi} \Lvac = \Phi_{\suN} \Lv_{\suN'}. 
\end{equation}

\subsection{Super vector spaces}
\label{Sec:Supvec}
  
A {\em super vector space} \cite{Var,Del} is $\ZZ_2$-graded vector space
\begin{equation}\label{supV}
 V = V_0 \oplus V_1,\quad 0,1 \in \ZZ_2 = \ZZ/2\ZZ.
\end{equation}
The complex super vector space is denoted $\CC^{d_0|d_1}$, 
where $d_i$ is dimension of $V_i$. 
The elements $v \in V_0$, $p(v) = 0$ and $w \in V_1$, $p(w) = 1$ are called {\em even} 
and {\em odd} respectively.

The $\ZZ_2$-graded (super) tensor product for such elements can be defined using {\em sign rule}
\begin{equation}\label{sox}
 v \sox u = (-1)^{p(v)\, p(u)} u \sox v.
\end{equation}

Roughly speaking, one $\suq$-qubit could be compared with element of $\CC^{1|1}$, 
but such approach encounters difficulties for more $\suq$-qubits.
Indeed, $\ZZ_2$-graded tensor product \Eq{sox} in definition 
\Eq{suq} should use {\em different} copies of initial space. 
Such approach may be quite natural in definition of $\ZZ_2$-graded tensor 
product of algebras and can be used for construction of 
Clifford algebras \cite{Post,spin3n}.

However, it is not quite clear, how to implement similar idea
for construction of $\suq$-qubits from $\CC^{1|1}$, because
implementation of $m$ {\em different} copies of $\suq$-qubit
may require to use bigger spaces such as $\CC^{m|m}$. 

Let us consider basis of $V=\CC^{m|m}$:
$e^0_k \in V_0$, $e^1_k \in V_1$, $k = 0,\ldots,m-1$. 
States of qubits $\alpha e^0_k + \beta e^1_k$ belong 
to {\em different} $2D$ subspaces of $V$ and their tensor product for 
$k=0,\ldots,m-1$ is the linear subspace with dimension $2^m$ of the `whole' 
tensor product $V^{\otimes m}$ with dimension $(2m)^m$.  
However, it may look as not very natural choice. 

For more trivial cases $\CC^{d_0|0}$ and $\CC^{0|d_1}$ the tensor
product of super vector spaces could be treated as symmetric and antisymmetric
tensors respectively, but in such a case all vector spaces in product are usually
considered as identical. Similar approach with identical copies of $\CC^{d_0|d_1}$ is
also quite common in supersymmetry. Thus, superspace is only briefly mentioned here
for comparison with other models of $\suq$-qubits and 
the term {\em super-indexed} is used earlier to emphasize the difference with known 
supersymmetric model of qubits \cite{squb}.

It should be mentioned also, that in the spinor model of $\suq$-qubits discussed in \Sec{Clif}
the elements $\Lv_\mu^a$ formally do not belong to superalgebra. Despite the 
super-commutation rule is valid for different super-indexes \Eq{LIcom}, 
it can be violated for the same one, \Eq{L0L1}.

\section{Conclusion}

Jordan-Wigner transformation maps some  operators \Eq{ladloc} acting `locally'
on $n$ qubits (or spin-$\frac{1}{2}$ systems) into $n$ fermionic creation and annihilation 
(ladder) operators. The `nonlocal' construction of such a map 
supposes introduction of some formal ordering on the set of qubits. 
Such ordering may be natural 
for some simple models such as 1D chain. All ladder operators in fermionic system are formally equivalent
and unnatural order produces technical difficulties for more general models such as multidimensional 
lattices and more general graphs. 

Antisymmetric algebra may be formally used for equal (unordered) description of ladder 
{\em operators}, but it does not answer a question about inequality of {\em states}. 
To address such a problem in this work was suggested model of `super-indexed' $\suq$-qubits.
Equivalence relation necessary for agreement with Jordan-Wigner transformation 
and anti-commutativity of ladder operators is {\em signed exchange rule\/} 
for $\suq$-qubits \Eq{supcom}.

Algebraic model of $\suq$-qubits with such property was also discussed. The model uses
Clifford algebras and spinors. Such approach is different with analogue constructions
more common in supersymmetric models only briefly discussed in subsection above.

\section*{Acknowledgements}

Author gratefully acknowledges possibility to present some topics considered here in a talk 
at Conference ``Quantum Informatics -- 2021,'' Moscow.

\end{document}